\documentclass[aps,prx,amsmath,amssymb,reprint,superscriptaddress]{revtex4-1}
\usepackage{graphicx}
\usepackage{amsmath}
\usepackage{dcolumn}
\usepackage{bm}
\usepackage{bbold}
\usepackage{color}
\usepackage{amsfonts}
\usepackage{amssymb}
\usepackage{mathrsfs}
\usepackage{tabularx}
\usepackage{braket}
\usepackage{mathtools}
\usepackage{soul}			

\usepackage{caption}
\usepackage{subcaption}

\begin{document} 
\title{Determination of the Anisotropy of Permittivity of Quantum Paraelectric Strontium Titanate}

\author{Maxim Goryachev}
\affiliation{ARC Centre of Excellence for Engineered Quantum Systems, School of Physics, University of Western Australia, 35 Stirling Highway, Crawley WA 6009, Australia}

\author{Lingfei Zhao}
\affiliation{Nanjing University, 22 Hankou Road, Gulou District, Nanjing, Jiangsu, China, 210093}

\author{Zijun Zhao}
\affiliation{Nanjing University, 22 Hankou Road, Gulou District, Nanjing, Jiangsu, China, 210093}

\author{Warrick G. Farr}
\affiliation{ARC Centre of Excellence for Engineered Quantum Systems, School of Physics, University of Western Australia, 35 Stirling Highway, Crawley WA 6009, Australia}

\author{Jerzy Krupka}
\affiliation{Instytut Mikroelektroniki i Optoelektroniki PW, Koszykowa 75, 00-662 Warsaw, Poland}

\author{Michael E. Tobar}
\email{michael.tobar@uwa.edu.au}
\affiliation{ARC Centre of Excellence for Engineered Quantum Systems, School of Physics, University of Western Australia, 35 Stirling Highway, Crawley WA 6009, Australia}

\date{\today}


\begin{abstract}

The dielectric properties of strontium titanate (SrTiO$_3$) have previously been reported from room temperature to low temperatures with conflicting results. In this work, precision measurement of the permittivity is undertaken by simultaneously measuring transverse electric and transverse magnetic resonant modes within a single crystal. It is unequivocally shown that the permittivity is isotropic at room temperature with a permittivity of order $316.3\pm2.2$ by measuring multiple modes of different electric field polarisations. As the crystal is cooled to 5 K and undergoes well known phase transitions, we show the material becomes uniaxial anisotropic with the ratio of the parallel to perpendicular permittivity to the cylinder z-axis of the sample as high as 2.4 below 6 K.

\end{abstract}

\maketitle

\section*{Introduction}
Strontium titanate (SrTiO$_3$) or STO has been a very popular material for study at low temperatures due to its unique properties qualifying this material as quantum paraelectric. The crystal is a perovskite at room temperatures and approaches the ferroelectric phase at low temperatures, however quantum fluctuations of low frequency phonon modes prevent it from undergoing the corresponding phase transition\cite{Muller:1979aa}. Since the discovery, there has been a number of attempts to study this phenomena at low temperatures both theoretically \cite{Yuan:2003aa,
Palova:2009aa} and experimentally \cite{Muller:1979aa, Neville:1972aa, Geyer:2005aa,Trainer:2001aa,Vendik:1998aa,Krupka:1994aa,Lytle:1964aa}. Moreover, the high permittivity and large dependence of permittivity on electric field below 70 K is an important characteristic, which allows device miniaturisation and small mode volume devices (a recent example includes room temperature masers \cite{Breeze:2015aa}) and the voltage control of superconducting electronic devices\cite{Lancaster:1998aa}. { Further utilisation of STO in different fields of low temperature physics extensive studies of its electromagnetic properties.} In this work, we characterise the permittivity of a crystalline STO sample and unequivocally show that the permittivity is isotropic at room temperature by measuring multiple modes of different electric field polarisations. As the crystal is cooled to 5 K and undergoes well known phase transitions, we show the material becomes uniaxial anisotropic. These results give important information for designing devices over this temperature range, which utilise the unique properties of STO.

Dielectric characterisation of materials using multiple modes in a single resonant structure has become a successful technique for uniaxial \cite{Tobar:1998aa,Krupka:1999aa,Krupka:1999ab} and biaxial permittivity determination\cite{Carvalho:2015aa}. These techniques can use a combination of low order Transverse Electric (TE) and Transverse Magnetic (TM) modes, as well as higher order modes such as Whispering Gallery Modes (WGM) to successfully characterise the anisotropy. The main advantages of this technique is due to the use of a large number of modes in the centimetre and millimetre wave frequency ranges with varying mode polarisation and high magnetic and electrical filling factors within the sample. Due to these properties WGM devices have also been implemented to characterise material electromagnetic loss and allow the comprehensive Electron Spin Resonance (ESR) spectroscopy of various ions within different crystals using the multiple modes\cite{PhysRevB.88.224426,quartzG,Goryachev:2014ab, Goryachev:2015aa, Creedon:2015aa,Creedon:2011wk}. 

\section*{Characterisation at Room Temperature}
At room temperature, it is well known that STO exhibits a cubic structure and therefore is most likely isotropic. Despite this, early work that implemented a capacitive technique using single crystal wafers between 0.125 and 1mm thickness in various crystal orientations, reported anisotropy at room temperature\cite{Neville:1972aa}. Gold electrodes were deposited on the samples and the permittivity was inferred from 4.2 K to 300 K and between 1 kHz to 50 MHz. They measured 330 in the [001] direction, 458 in the [011] direction and 448 in the [111] direction at room temperature. This type of scatter in measurements can be evident with only a -0.02 mm systematic error in thickness measurement. In the next section we compare our results across the whole temperature range and discuss the most likely discrepancies between our measurements and the ones presented in Ref\cite{Neville:1972aa}.

\begin{figure}[ht!]
	\centering
			\includegraphics[width=0.46\textwidth]{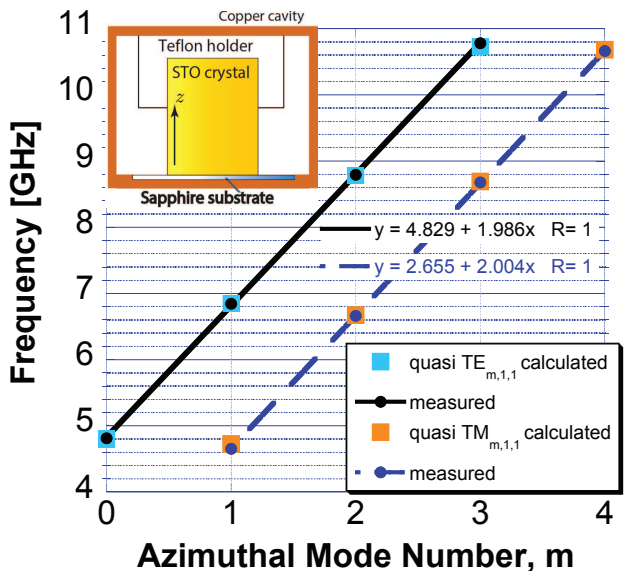}
	\caption{Comparison of measured and calculated frequencies of the fundamental quasi-TE and TM mode families with the a STO permittivity of $316.3$. Inset: Side view or $(r,z)$ plane of the STO cylindrical cavity. The cylindrical crystal is mounted on a sapphire disc substrate and held in place by a teflon piece.}
	\label{expset}
\end{figure}

In order to probe the dielectric properties of STO, a cylindrical crystal specimen of diameter 3.27 mm and height 3.66 mm was utilised. The crystal was put in an cylindrically symmetric oxygen free copper cavity on a sapphire substrate and kept in place from above by a teflon piece as demonstrated in Fig.~\ref{expset}. { The sapphire substrate and the teflon piece are utilised in order uncouple the crystal modes from the cavity walls and, thus, to increase the Quality Factor. On the other hand, these materials do not introduce any significant error to the measurement results due to orders of magnitude difference in permittivity with the STO crystal. Sapphire is chosen on the bottom of the cavity due to it low loss and good thermal conductivity at low temperatures.} Microwave radiation in the cavity was excited via both straight electrode antenna and loop probes in order to resolve modes of TE and TM polarisations. The system is characterised in transmission with a vector network analyser. \\

To calculate mode frequencies from comparison with experiment, finite element software was implemented {to simulate the resonator in Fig.~\ref{expset}. This model utilised known parameters of all materials except for the STO crystal. For the considered structure,} only modes with $m=0$ can be pure TM or TE polarisation, for $m>1$ the modes are quasi TE and TM and become WGM as $m$ increases. The first four modes of the fundamental quasi TE and TM mode families were measured with azimuthal mode numbers of up to $m = 4$. The permittivity was iterated to minimise the error between the calculated and measured frequencies. Fig.~\ref{expset} shows clearly good agreement, with the value $316.3$ minimising the aggregate error between simulation and experiment. The cylindrically symmetric permittivity gives the average error of the four TE modes to be $-0.18\%$ while the average of the four TM modes to be $0.52\%$. Also, dimensions were measured to the nearest hundredth of a mm using laser technique giving approximately $0.3\%$ error in height and radius. Adding all errors in quadrature gives a value of permittivity of $316.3\pm2.2$, consistent with more recent measurements of permittivity, which have only implemented TE modes\cite{Krupka:1994aa,Breeze:2015aa}. The mode free-spectral range is nearly constant and close to 2 GHz and the permittivity was demonstrated to be effectively constant and shows no frequency dependence from 4 to 11 GHz. This result is consistent with the soft mode phonon theory, which dominates the value of permittivity in this frequency range\cite{Neville:1972aa}. Prior work predicts the permittivity to be constant beyond a THz at room temperature and beyond 100 GHz at 4.2 GHz\cite{Neville:1972aa}.

\cite{Lytle:1964aa}

\begin{figure}[ht!]
	\centering
			\includegraphics[width=0.45\textwidth]{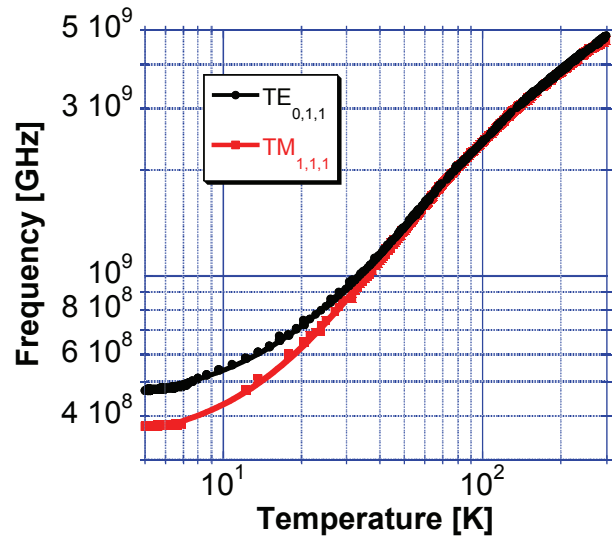}
	\caption{Frequency versus temperature for the TE$_{011}$ mode and the quasi TM$_{111}$ mode. Due to the high permittivity electrical energy filling factors sum up close to unity (or 100\%) across the whole temperature range. The circle and square dots represents measurements of frequency, while the line represents a polynomial fit.}
	\label{freq}
\end{figure}

\section*{Revealing Crystal Anisotropy at Cryogenic Temperatures}
For cryogenic measurements, the copper cavity was attached to the 4K plate of a cryogen free pulsed tube system and measured in transmission using a Vector Network Analyzer. In order to measure the temperature dependence, a Cernox calibrated thermometer was  attached to the copper cavity and continuously measured throughout the experiment. Measurements were implemented by warming up and cooling down very slowly in equilibrium, with temperature lag between the thermometer and crystal kept to a minimum. The frequency versus temperature measurements are shown in Fig.~\ref{freq} of the TE$_{011}$ mode and the quasi TM$_{111}$ mode. Due to the high dielectric constant the electrical energy filling factors are approximately constant across the temperature range, for the TM$_{111}$ mode the electrical energy filling factor parallel to the c-axis is $Pe_{||}\approx 63\%$ and perpendicular is $Pe_{\bot}\approx37\%$, while for the TE$_{011}$ mode it is $Pe_{||}= 0\%$ and $Pe_{\bot}\approx100\%$. Thus, previous measurements that utilised TE modes of $m=0$ only determine the perpendicular component of permittivity\cite{Krupka:1994aa,Geyer:2005aa}. This is overcome by simultaneously measuring the TM$_{111}$ mode, which couples to both anisotropy directions, but more strongly to the parallel component. Thus, through numerical iteration both components of permittivity can be found. Due to the non-zero value of $m$ for the TM$_{111}$ mode, the resonance is in principle degenerate. The doublet degeneracy will be lifted if the permittivity is biaxial, and this effect has recently been used to characterise biaxial material\cite{Carvalho:2015aa}. {Since this effect is not observed for STO at any temperature,} we can conclude the crystal does not exhibit biaxial permittivity. However, it is clear from the temperature dependence observed of the mode frequencies as shown in Fig.~\ref{freq}, that the permittivity exhibits uniaxial anisotropy at low temperatures.

\begin{figure}[ht!]
	\centering
			\includegraphics[width=0.4\textwidth]{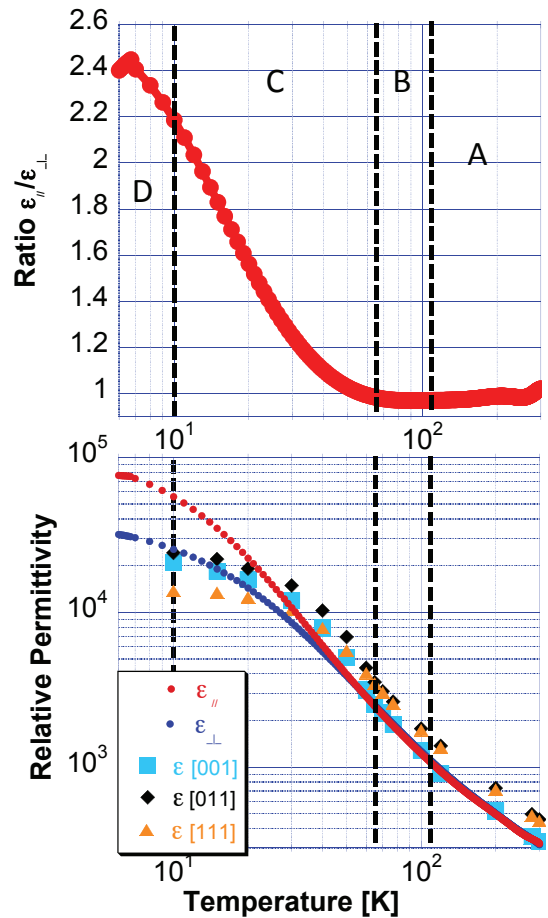}
	\caption{Above: Anisotropy ratio $\frac{ \epsilon_{||}}{\epsilon_{\bot} }$ for STO. Below: Derived permittivities $\epsilon_{||}$ and $\epsilon_{\bot}$ from the frequency measurement in Fig.~\ref{freq}, with comparison to measurements from Ref\cite{Neville:1972aa} along different crystallographic directions}
	\label{anisotropy}
\end{figure}

The values of permittivity derived from the curve fits in Fig.~\ref{freq} and the finite element modelling are shown in Fig.~\ref{anisotropy} are compared with data from\cite{Neville:1972aa}. {All these results show significant change in anisotropy that has been explained by crystal structure phase transitions. Plots in Fig.~\ref{anisotropy} are subdivided into sections A-D in accordance with corresponding phases.
From room temperature down to 110K,} STO exhibits highly symmetric cubic perovskite structure (space group Pm3m) as demonstrated by section A. For lower temperatures, the crystal structure is deformed from cubic by a number of effects. In the range $110-65$K (see section B in Fig.~\ref{anisotropy}), STO demonstrates tetragonal (space group I4/mcm) structure shifting to orthorhombic in the range $55-35$K (section C). Also, it has been demonstrated that at $10$K, STO forms a single low-symmetry phase (section D), which is possibly a rhombohedral structure\cite{Cao:2000aa}, \cite{Lytle:1964aa}. It is evident from Fig.~\ref{anisotropy} that the main transition to biaxial anisotropy occurs when the crystal structure shifts to the orthorhombic phase.

The results from ref\cite{Neville:1972aa} do not show any clear anisotropy as originally claimed, and the varying results are most likely due to systematic scatter of data due to dimension errors. Also, below 50 K the results of\cite{Neville:1972aa} are not clearly anisotropic either, and in general give a lower value of permittivity below 15 K than those measured in this work. It has been shown that the application of stress to the crystal lowers the permittivity of the crystal \cite{Muller:1979aa}, and that the permittivity of STO is highly stress dependent. We postulate that the results of \cite{Neville:1972aa} had excess stress created in the crystal due to the differential contraction of the gold electrodes deposited on the crystal. Our technique is stress free as the crystal sits on a sapphire substrate, with an indentation slightly bigger than the crystal radius to keep the crystal in place, along with a loosely fitting teflon cap held from the top of the cavity. 

In summary, we have undertaken precision measurement of the permittivity of Strontium Titanate from room temperature to 5 K. We unequivocally showed by using multiple modes that the permittivity is isotropic at room temperature, consistent with its well known cubic structure at room temperature. We have adapted the WGM technique to measure the anisotropy at cryogenic temperatures and shown the crystal becomes biaxial anisotropic when it undergoes a phase transition to a rhombohedral structure.

\section*{Acknowledgements}
We thank Prof. Andrew Johnson for useful discussions regarding this work. This work was supported by the Australian Research Council Grant No. CE110001013.

\section*{References}

%

\end{document}